\documentclass[aps,pra,twocolumn,amsmath,amssymb,superscriptaddress]{revtex4-2}

\usepackage{graphicx}
\usepackage[tight]{units} 
\usepackage{textcomp} 
\usepackage{gensymb} 
\usepackage{bbold}
\usepackage{dsfont}
\usepackage{siunitx}
\usepackage[normalem]{ulem} 

\usepackage{color}

\begin{document}

\title{Controlling long ion strings for quantum simulation and precision measurements}

\author{Florian Kranzl}
\affiliation{Institut f\"ur Quantenoptik und Quanteninformation, \"Osterreichische Akademie der Wissenschaften,
Technikerstra\ss{}e 21a, 6020 Innsbruck, Austria}
\affiliation{Institut f\"ur Experimentalphysik, Universit\"at Innsbruck, Technikerstra\ss{}e 25, 6020 Innsbruck, Austria}

\author{Manoj K. Joshi}
\affiliation{Institut f\"ur Quantenoptik und Quanteninformation, \"Osterreichische Akademie der Wissenschaften,
Technikerstra\ss{}e 21a, 6020 Innsbruck, Austria}

\author{Christine Maier}
\altaffiliation[Current address: ]{AQT, Technikerstra{\ss}e 17, 6020 Innsbruck, Austria}
\affiliation{Institut f\"ur Quantenoptik und Quanteninformation, \"Osterreichische Akademie der Wissenschaften,
Technikerstra\ss{}e 21a, 6020 Innsbruck, Austria}
\affiliation{Institut f\"ur Experimentalphysik, Universit\"at Innsbruck, Technikerstra\ss{}e 25, 6020 Innsbruck, Austria}

\author{Tiff~Brydges}
\altaffiliation[Current address: ]{Department of Applied Physics, University of Geneva, Geneva, Switzerland}
\affiliation{Institut f\"ur Quantenoptik und Quanteninformation, \"Osterreichische Akademie der Wissenschaften,
Technikerstra\ss{}e 21a, 6020 Innsbruck, Austria}
\affiliation{Institut f\"ur Experimentalphysik, Universit\"at Innsbruck, Technikerstra\ss{}e 25, 6020 Innsbruck, Austria}

\author{Johannes Franke}
\affiliation{Institut f\"ur Experimentalphysik, Universit\"at Innsbruck, Technikerstra\ss{}e 25, 6020 Innsbruck, Austria}

\author{Rainer Blatt}
\affiliation{Institut f\"ur Quantenoptik und Quanteninformation, \"Osterreichische Akademie der Wissenschaften,
Technikerstra\ss{}e 21a, 6020 Innsbruck, Austria}
\affiliation{Institut f\"ur Experimentalphysik, Universit\"at Innsbruck, Technikerstra\ss{}e 25, 6020 Innsbruck, Austria}

\author{Christian F. Roos}
\affiliation{Institut f\"ur Quantenoptik und Quanteninformation, \"Osterreichische Akademie der Wissenschaften,
Technikerstra\ss{}e 21a, 6020 Innsbruck, Austria}
\affiliation{Institut f\"ur Experimentalphysik, Universit\"at Innsbruck, Technikerstra\ss{}e 25, 6020 Innsbruck, Austria}
\email{christian.roos@uibk.ac.at}

\date{\today}

\begin{abstract}
Scaling a trapped-ion based quantum simulator to a large number of ions creates a fully-controllable quantum system that becomes inaccessible to numerical methods. When highly anisotropic trapping potentials are used to confine the ions in the form of a long linear string, several challenges have to be overcome to achieve high-fidelity coherent control of a quantum system extending over hundreds of micrometers. In this paper, we describe a setup for carrying out many-ion quantum simulations including single-ion coherent control that we use for demonstrating entanglement in 50-ion strings. Furthermore, we present a set of experimental techniques probing ion-qubits by Ramsey and Carr-Purcell-Meiboom-Gill (CPMG) pulse sequences that enable detection (and compensation) of power-line-synchronous magnetic-field variations, measurement of path length fluctuations, and of the wavefronts of elliptical laser beams coupling to the ion string.
\end{abstract}

\maketitle

\section{\label{sec:Introduction} Introduction} 
Trapped ions have been one of the leading candidates for precision and metrology research due to their high degree of isolation from the surrounding perturbations \cite{wolf2019motional,ludlow2015optical}.
More recently, the development of entangling operations in qubits encoded in trapped ions \cite{Ballance:2016,Clark:2021,Srinivas:2021} has led to high fidelity quantum computation and simulations 
\cite{Bermudez:2017,Bruzewicz:2019,Monroe:2021}
In order to fully utilize the power of quantum computation and simulations, one needs quantum systems comprised of a large number of (quantum-error corrected) qubits. There has been a strong push to scale up the number of trapped ions by designing specific trap geometries for shuttling and reconfiguring ion crystals \cite{Zak:2020,Pino:2021} or arrays of ion traps \cite{Hakelberg:2019,Jain:2020}, and by potentially connecting them via quantum networks \cite{kimble2008quantum}. In order to trap large-sized crystals, Penning traps, which utilize static electric and magnetic fields to trap charged particles \cite{Mavadia_2013, Britton:2012}, have been used for trapping planar crystals of several hundred ions and first quantum simulations have been carried out \cite{Bohnet2016}. However, due to the rotation of ion crystals in a Penning trap, individual qubit control, which is an essential ingredient for developing fully programmable quantum simulators \cite{Monroe:2021, altman2021quantum}, has not been demonstrated yet.    

Controlling the quantum state of a many-ion crystal presents various challenges: (1) The large number of ions implies a large number of motional modes that need to be cooled to low temperatures \cite{jordan2019near, lechner2016electromagnetically}. The collision rate of background gas molecules with the ions increases in proportion to the number of ions, which can give rise to frequent melting events of the crystal that require recrystallization. (2) Many-ion linear crystals require highly anisotropic trapping potentials which are typically achieved by lowering the confinement along the axis of the ion string, resulting in  high motional occupation numbers after Doppler cooling and a high heating rate of the center-of-mass motion in the direction of the ion string. (3) The large spatial extent of many-ion strings results in a spread of qubit transition frequencies because of spatially varying electromagnetic fields, and (4) it makes single-ion addressing across the entire ion string difficult to achieve.

\begin{figure}
    \centering
    \includegraphics[width=0.45\textwidth]{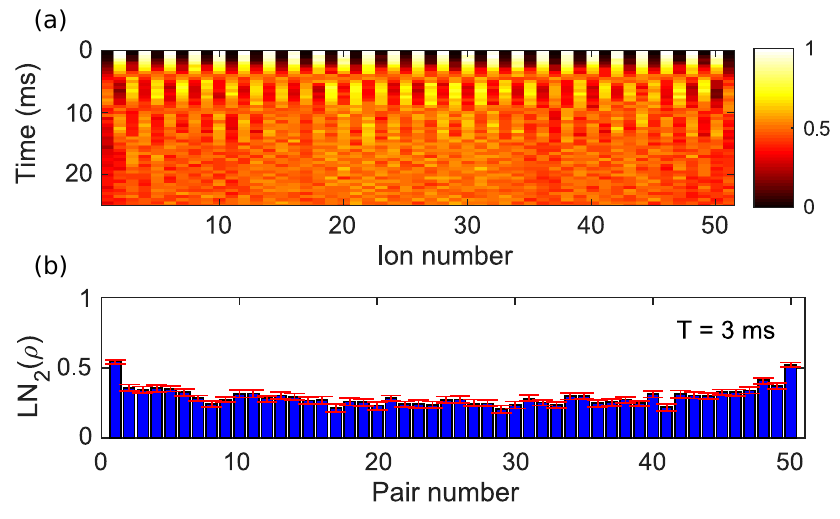}
    \caption{(a) Magnetization dynamics under a XY-Hamiltonian for a 51-ion chain initially prepared in a N\'eel state. (b) Log-negativity measured for connected pairs after $T=\SI{3}{ms}$ of time evolution.}
    \label{fig:EntaglementGen}
\end{figure}

In this manuscript, we present a set of techniques for controlling and characterizing long crystals containing tens of ions.  We extended single-ion addressing to strings of up to 51 ions which enable us to investigate many-body non-equilibrium dynamics with tens of ions and to demonstrate entanglement between all pairs of 
ing ions (see Fig.~\ref{fig:EntaglementGen}) as discussed later in more detail. The manuscript is structured as follows: In section \ref{sec:Trap}, we describe the experimental platform and discuss sub-Doppler cooling techniques for long ion strings. We discuss the collision rate of the background gas with the ion crystal which limits the time available for performing experiments with the ions. In section \ref{sec:coherence}, we describe techniques for characterizing the electromagnetic environment of the ion crystal and discuss how to mitigate errors in coherent control of long ion strings. In particular, we present two applications of probing external fields with the ion crystal using a CPMG sequence: Firstly, CPMG sequences exciting the ion-qubits are used for probing the line-synchronous components of the magnetic field, which enable a feedforward compensation of these disturbances. Secondly, CMPG sequences are employed for characterizing the wavefronts of a laser beam that collectively couples to all qubits; by these measurements, the laser beam direction can be optimized and its wavefront curvature at the location of the ions can be reduced. Furthermore, we show Ramsey experiments done with beams from two separate light paths which we use to assess path-length fluctuations in the optical fibers delivering light to the ions. In section \ref{sec:Addressing}, we show the results of the single-ion addressing of the full 51-ion crystal. In section \ref{sec:entanglement}, entanglement generation in the 51-ion crystal is demonstrated by a measurement of the logarithmic negativity for connected pairs and triplets.

\section{\label{sec:Trap} Trap characterization} 

\subsection{\label{sec:setup}{Experimental setup}} 
\begin{figure}
\centering
\includegraphics[width=0.45\textwidth]{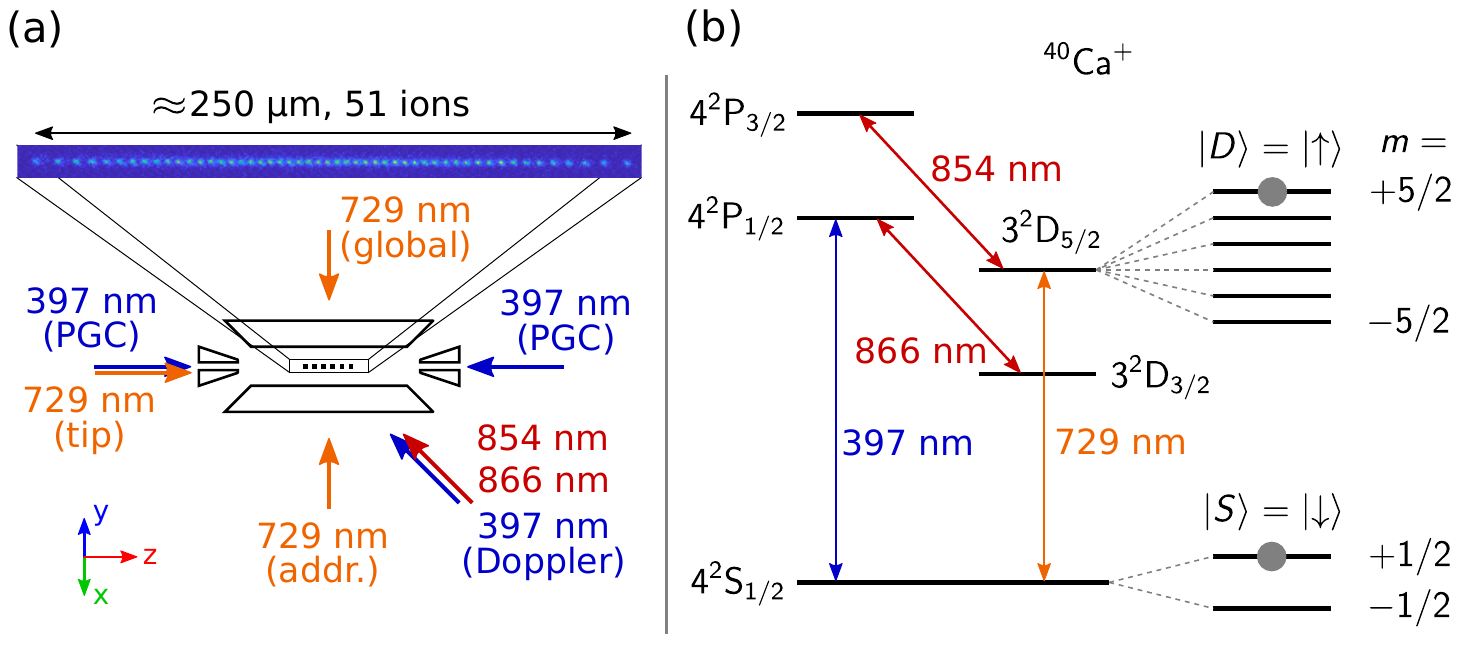}
\caption{(a) Ion string in a linear Paul trap. A string of up to 51 ions is cooled, manipulated, and detected with laser beams impinging from various angles. Two counter-propagating beams creating a polarization gradient along the axis of the ion string are directed through holes in the ion trap's endcaps. (b) Partial level scheme of $^{40}$Ca$^+$. The qubit is encoded in the states $4^2\mathrm{S}_{1/2} (m=+1/2)$ and  $4^2\mathrm{D}_{5/2} (m=+5/2)$ (grey dots).}
\label{fig:level_scheme_setup}
\end{figure}

The experimental platform discussed in this manuscript has been developed for quantum simulations with trapped atomic ions. The centerpiece of the experiment is a linear Paul trap, used for confining long strings of $^{40}$Ca$^+$ ions (Fig.~\ref{fig:level_scheme_setup}). The trap potential is made highly anisotropic by keeping the confinement along the trap's symmetry axis (along the `z' axis) fairly weak to allow for confining up to 51 ions as a linear string. The calcium ions are created in the trapping region by photo-ionizing neutral calcium atoms emanating from a resistively heated oven that contains pure calcium in solid form. The ions are captured by the trapping fields and cooled via Doppler cooling. The Doppler cooling laser beam propagates at 45 degrees to the trap axis, while having roughly equal overlap with the two other principal trap axes (`x' and `y'). The confinement in the radial plane (`xy' plane) is made slightly anisotropic to enable efficient Doppler cooling of all motional modes. Typical values for the radial trapping frequencies are $\omega_x \approx 2\pi\times \SI{2.93}{MHz}$ and $\omega_y \approx 2\pi\times \SI{2.89}{MHz}$ and for the axial trapping frequency $\omega_z \approx 2\pi \times \SI{127}{kHz}$. A set of SmCo magnets defines the quantization axis in the trapping region by creating a magnetic field of $B=\SI{4.17}{gauss}$ pointing along the trap axis. Further details can be found in ref.~\cite{Hempel2014}. 

A typical experimental sequence to coherently control the electronic and motional states (Fig.~\ref{fig:level_scheme_setup}) consists of the following steps: As a first step, ions are cooled approximately to the Doppler temperature by a \SI{397}{nm} laser beam, which is red-detuned by about half a linewidth from the S${}_{1/2}\leftrightarrow$ P${}_{1/2}$ transition, in conjunction with lasers at \SI{866}{nm} and \SI{854}{nm}  pumping out the metastable D$_{3/2}$ and D$_{5/2}$ levels. The Doppler cooling duration is set to \SI{3}{ms}. The second step, sub-Doppler cooling of ions, which is essential to achieve high fidelity quantum control, is carried out by polarization-gradient and resolved sideband cooling techniques. During sideband cooling, a laser at \SI{854}{nm} pumps out population from the D$_{5/2}$ level. For further details, see section \ref{sec:Cooling}. As a third step, the coherent manipulation of ions is performed on the S${}_{1/2}$ $\leftrightarrow$ D$_{5/2}$ transition with a laser at \SI{729}{nm} having a linewidth of less than 10 Hz. In this step, ions are prepared in coherent superpositions of their internal states or entangled with each other.
In the final step, the state of the individual ions is detected by imaging the fluorescence emitted by the ions onto an EMCCD camera.

The key ingredients of quantum simulation experiments are single-qubit operations, with individual and global controls, and entangling operations. The laser beams at \SI{729}{nm} performing these operations are sketched in Fig.~\ref{fig:level_scheme_setup}(a). The entangling interaction is generated by a bichromatic laser beam with frequencies $\omega_\pm$, that simultaneously interacts with the motional and electronic degrees of freedom of all trapped ions. The laser beam illuminates the ion chain from a radial direction such that it couples to the $2N$ transverse motional modes, which mediate a M{\o}lmer-S{\o}rensen interaction between the qubits. In this way, we achieve a variable long-range interaction \cite{Porras:2004}. The interaction Hamiltonian is expressed as  
\begin{equation}
H=\sum_{i < j} J_{ij} \sigma_i^x\sigma_j^x + B\sum  _k \sigma_k^z,
\end{equation}
where $\sigma_k^\alpha$ ($\alpha =x,y$ and $z$) are Pauli operators for ion index $k$. The elements of the spin-spin matrix ($J_{ij}$) are set by the bichromatic laser beam intensity at the ions and the laser detuning from the sidebands; the strength of the transverse field $B$ is controlled by the centerline detuning $\delta=(\omega_+-\omega_-)/2-\omega_0$, where $\omega_0$ is the qubit transition frequency \cite{Jurcevic:2014}. The $J_{ij}$ matrix and $B$-field terms are related to the laser-ion interaction parameters via
\begin{equation}
J_{ij}=\frac{\Omega_i \Omega_j}{2}\sum_{m=1}^{2N} \frac{\eta_{i,m}\eta_{j,m}}{\Delta_m},\quad
B=\delta/2,
\end{equation}  
where $\Omega_i$ is the Rabi frequency of the frequency components of the bichromat beam coupling to ion $i$, $\eta_{i,m}$ is the Lamb-Dicke parameter of ion $i$ and mode $m$, $\Delta_m$ is the detuning of bichromatic beam from the motional mode of interest.

\subsection{\label{sec:Cooling} Laser cooling and motional heating of long strings}

Sub-Doppler cooling of long ion strings is a prerequisite for most entanglement-generating atom-light interactions employed in quantum information processing, quantum simulation, and quantum metrology. The challenge is to efficiently cool many modes (possibly spread over hundreds of kilohertz) close to the motional ground state, in the shortest possible time.
All radial modes need to be ground-state cooled in order to reduce fluctuations in coupling strength and AC-Stark shifts for the realization of single- and multi-qubit gate operations with high fidelity. In our experiments, ground-state cooling of all $2N$ radial modes of an $N$-ion string is achieved by resolved sideband cooling.

\begin{figure}
\centering
\includegraphics[width=0.45\textwidth]{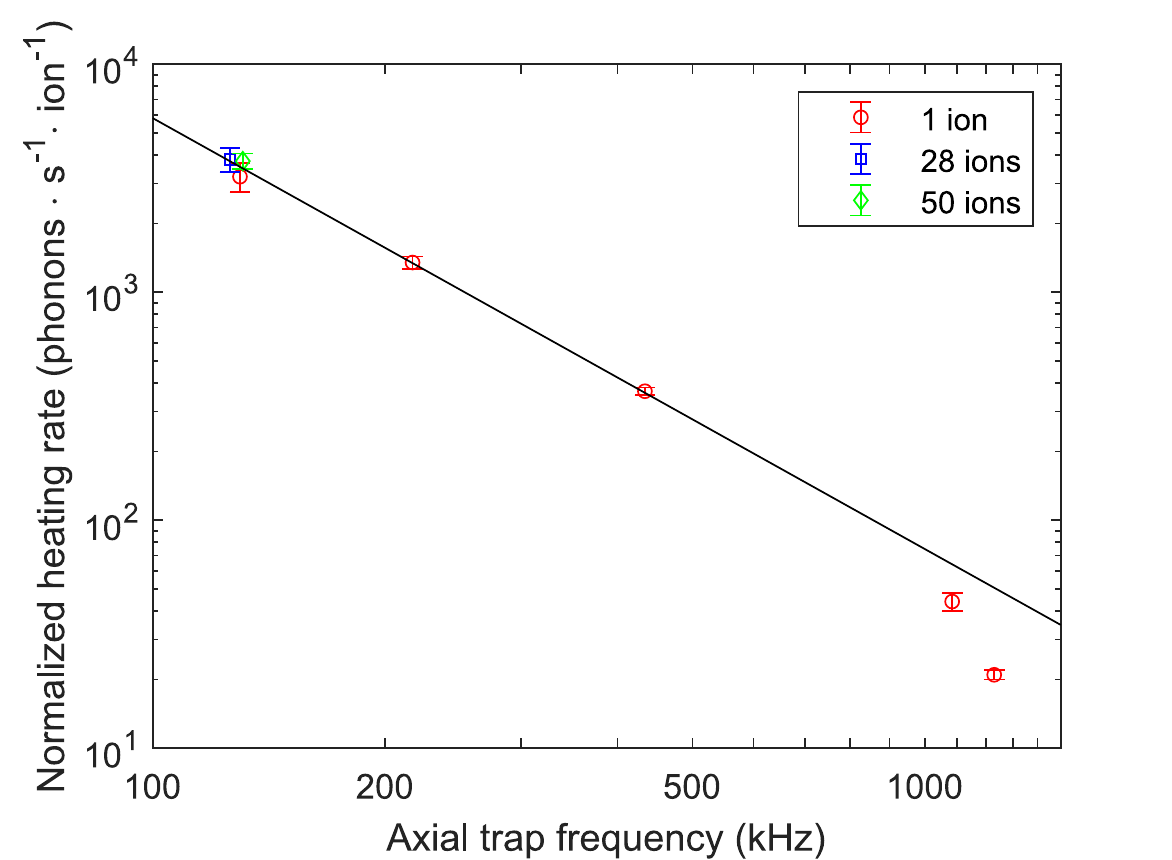}
\caption{Normalized heating rate vs. axial trap frequency, normalized by the number of ions $N$. Open circles show data measured with a single ion (red), 28 ions (blue) and 50 ions (green). The red line is a fit of an expression $dn/dt \propto  \frac{1}{\omega^{\alpha}}$ to the experimental data, where we estimate $\alpha = 1.9(2)$.}
\label{fig:HeatingRates}
\end{figure}

Due to weak confinement along the principal trap axis, the longitudinal motion suffers from a considerable motional heating \cite{brownnutt2015ion} which, in combination with the weak confinement, gives rise to a rather weak localization of the ions as compared to the laser wavelength. This compromises the quality of single-qubit gate operations for long experimental sequences, because high phonon numbers result in a large spatial spread of the individual ions such that the laser-ion coupling strength of the addressing beam decreases with increasing size of the  ions’ wavepacket (see Fig.~\ref{fig:Addr_vs_time} in Section~\ref{sec:Addressing}). Therefore, we cool the longitudinal motional modes before and after sideband cooling of the radial modes, via polarization-gradient cooling~\cite{Joshi:2020}. Fig.~\ref{fig:HeatingRates} shows the longitudinal COM mode's heating rate normalized to the number $N$ of ions, $\frac{dn/dt}{N}$, as a function of the trapping frequency $\omega$ for $N=1$, $N=28$ and $N=50$ ions. The measurements show an increase of the heating rate in proportion to the number of ions and a dependence on the trapping frequency $\omega$ as $dn/dt \propto  \frac{1}{\omega^{\alpha}}$, where $\alpha = 1.9(2)$.

\subsection{\label{sec:Collision rates} Langevin collisions and recrystallization}
Laser-cooled atomic ions stored in an ultrahigh vacuum still experience Langevin collisions with background gas molecules. These collisions heat the ions and can even melt the ion Coulomb crystals or, less frequently, turn atomic ions into unwanted molecular ions \cite{obvsil2019room, hankin2019systematic}. 
As the rate at which Langevin collisions occur increases in proportion with the number of trapped ions, efficient recrystallization routines are required in order to maximize the time the ion crystal is available for a quantum physics experiment. 

\begin{figure}
\centering
	\includegraphics[width=0.45\textwidth]{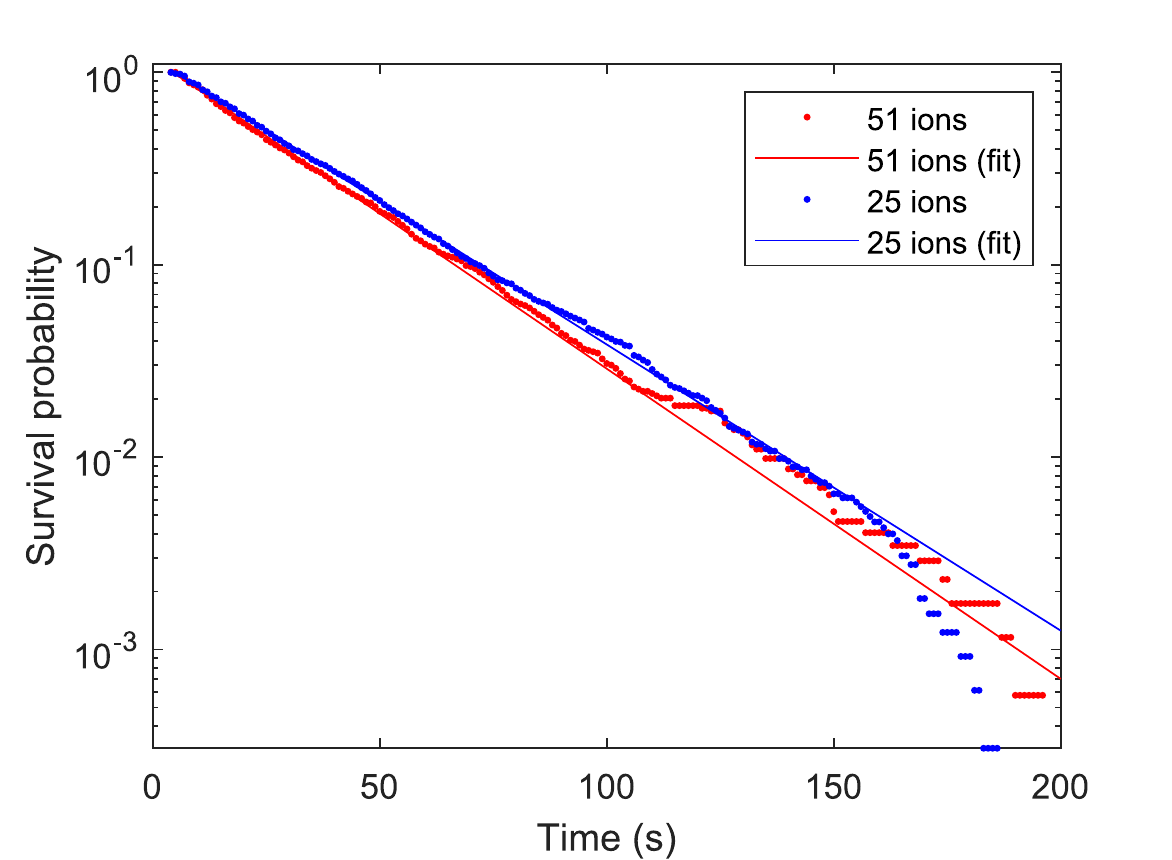}
	\caption{Survival probability of the ion crystal against melting by collisions vs. time for 25-ion and 51-ion strings.}
	\label{fig:RefreezeRate}
\end{figure}

We detect collision-induced melting of the ion crystals that occurs prior to the start of an experimental sequence by a drop of the number of photons detected while Doppler cooling the ions. When a collision is detected, the melting event is time-tagged and a refreeze sequence is called. In a refreeze sequence, the ions are cooled by a cooling beam that is red-detuned by $220$ MHz from the $\mathrm{S}_{1/2} \leftrightarrow \mathrm{P}_{1/2} $ atomic transition and with an intensity that is about 50\% higher than the one of the Doppler-cooling beam. The recooling beam's detuning was empirically optimized to quickly restore the crystalline order in 50-ion strings. In Fig.~\ref{fig:RefreezeRate}, we show the survival probabilities for 25- and 51-ion crystals as a function of time. Both ion crystals are studied at $\omega_z = 2\pi \times 127$ kHz and $\omega_x = 2\pi \times 2.93$ MHz. A fit of the curves with an exponentially decaying function yields life times of $\tau_s$= 29.2(2) s and 27.0(2) s for the 25- and 51-ion crystals, respectively. Even though the collision rate should be twice as high for the 51-ion crystal, both crystals exhibit nearly identical melting rates. Clearly, some Langevin collisions transfer an amount of energy to an ion that is not high enough to melt the ion crystal \cite{Pagano:2018}. It seems that for larger ion crystals, the redistribution of motional energy among all ions is quick enough to reduce the deleterious effect of radio-frequency heating such that Doppler cooling with standard parameters suffices to recool the ions quickly.

In addition to melting of ion crystals, we also experimentally detect collision events which leave the ions in a high mean phonon state. Most importantly, collisions that occur during sideband cooling or coherent probing leave the ions so hot as to affect the quantum operation. In this case, the ions are mostly detected as being in the bright state, because the energy transferred to the crystal is still too small to have melted the ion string, yet high enough to strongly reduce the excitation probability on the quadrupole transition. We detect these events by applying a resonant carrier $\pi$-pulse during the probe time. For a 51-ion crystal, we measure a total rate of $7(3)\times10^{-5}$ collisions/ms.  

We also detect collisions turning calcium ions into molecular ions by the appearance of non-fluorescing ions in the camera images. Heavier molecular ions, most likely CaOH$^+$ are deterministically eliminated by lowering the rf-trapping potential. In some events, we observe formation of CaH$^+$ molecules \cite{kimura2011sympathetic},  which has a mass very close to the calcium ion itself, making it difficult to eliminate the ion from the trap. This molecule has a dissociation channel for blue light and hence is dissociable via light at \SI{375}{nm}, used for photoionization of calcium atoms in our system or the Doppler cooling (\SI{397}{nm}) laser beam  \cite{khanyile2015observation}. We observe about six ions turning dark per 24 hours for a 51-ion string and on an average two to three ions for a 25-ion string.

\section{\label{sec:coherence} Coherent manipulation of long qubit registers} 

\subsection{\label{sec:LineCycle} Compensation of line-synchronous magnetic field fluctuations}
Power supplies and other electrical devices contribute to ambient magnetic field in the laboratory. The $50\,$Hz electricity mains (AC power line) especially, powering the laboratory devices, induces a line-synchronous alternating magnetic field. Nonlinearities in the electrical loads generate components with higher harmonics of 50$\,$Hz, with odd-multiple components generally being stronger than the even ones. This ambient magnetic field leads to a modulation of the ions' transition frequencies. A mu-metal shield surrounding the vacuum system reduces these fluctuations; however, it does not fully protect the qubits from the fluctuating magnetic fields that arise from some power cables required to run  inside the mu-metal shield. In addition to the shielding we use a line trigger system, in which the experiment is synchronized with the $50\,$Hz power line cycle. In this way, the experiment is in phase with the line-synchronous magnetic field and the ion experiences the same modulation of the transition frequencies in every experiment. The line trigger eliminates dephasing due to line-synchronous magnetic field fluctuations, at the expense of a deterministic time-variation of the laser-ion detuning. To suppress these temporal variations, we built a feedforward system that applies a time-varying current to a pair of  magnetic field coils such that the generated field cancels out temporal variations of the ambient magnetic field at the location of the ions
\cite{merkel2019magnetic}.

\begin{figure}
    \centering
    \includegraphics[width=0.48\textwidth]{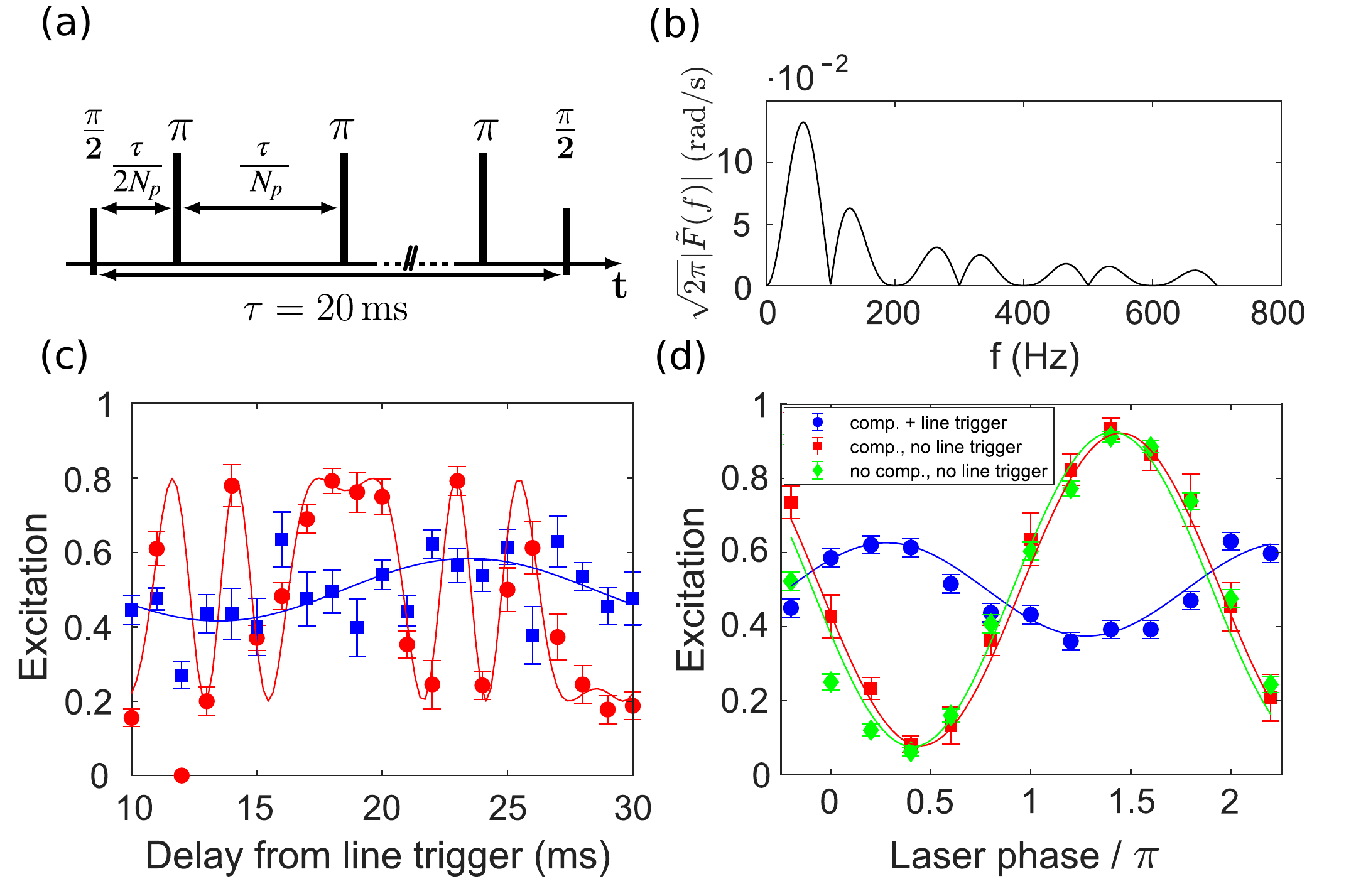}
    \caption{(a) Pulse timing of a CPMG pulse sequence containing $N_p$ $\pi$-pulses and total duration $\tau$, (b) Magnitude of a filter function
    with $N_p=2$ and $\tau=20\,$ms. (c) Mean excitation after a CPMG sequence ($N_p=2$, $\tau=20$~ms) probing 50~Hz noise.
    before (red circles) and after (blue squares) field compensation.
    Solid curves are fits of Eq.~\eqref{eq:CPMG_fit} to the data. (d)~Ramsey measurement with $\tau=4.5\,$ms probe time.     }
    \label{fig:CPMG_seq_mag_noise}
\end{figure}

For detecting line-synchronous magnetic field contributions at a particular frequency, we employ a $\pi$-pulse sequence \cite{kotler2013nonlinear}, namely a Carr-Purcell-Meiboom-Gill (CPMG) sequence~\cite{cpmg}. As illustrated in Fig.~\ref{fig:CPMG_seq_mag_noise}(a) the time separation between the two embedding $\frac{\pi}{2}$-pulses and the first (last) $\pi$-pulse is half the interpulse separation. A general $\pi$-pulse sequence acts as a filter with the filter function in the frequency domain given by ~\cite{biercuk2009experimental}
\begin{equation}
    \tilde{\mathcal{F}}(f)= \frac{1}{\sqrt{2\pi}i\omega}\left[1+(-1)^{N_p+1}e^{i\omega\tau}+2\sum_{j=1}^{N_p} (-1)^je^{i\omega\tau\delta_j}\right]\,
    \label{eq:filter_func}
\end{equation}
with the overall sequence duration $\tau$, the angular frequency $\omega=2\pi f$, the number of $\pi$-pulses $N_p$ and the occurrence of the $j^{th}$ pulse at $\tau\delta_j$. In the case of a CPMG sequence, $\delta_j=\frac{j-\frac{1}{2}}{N_p}$. In this formula, the length of $\pi$-pulses is considered to be negligible, which is a good approximation in our experiment, where the length of a $\pi$-pulse is on the order of $\approx 5\,\mu$s compared to the sequence length $\tau=20\,$ms. The absolute value of a filter function with two $\pi$-pulses is shown in Fig.~\ref{fig:CPMG_seq_mag_noise}(b). From the graph, one can see that the maximum of the filter function occurs at a frequency $f_{max}\approx\frac{N_p}{2\tau}$. Smaller local maxima occur for odd integer multiples of $f_\mathrm{max}$. Signals with frequencies $f=\frac{k}{\tau}$ for even numbers of pulses and $f=\frac{k}{\tau}+\frac{1}{2\tau}$ for odd numbers of pulses are suppressed, except $f_\mathrm{max}$ and its odd multiples. This feature enables us to probe the different frequency components of line-synchronous noise one by one. To characterize the modulation amplitude $A$ (in rad/s) and phase $\phi_\mathrm{noise}$, we scan the start $t_0$ of the CPMG sequence with respect to the line trigger event over one period of the probed noise component. If only a single noise component at frequency $f$ is present, one expects the excited state probability to be given by
\begin{align}
    P_{\uparrow}(A,t_0)=&\frac{1}{2}+\frac{C}{2}\sin\bigg(\sqrt{2\pi}|\tilde{\mathcal{F}}(f;N_p,\tau)|\nonumber\\&\cdot A\sin\Big(2\pi f t_0 +\phi_\mathrm{noise}+\arg(\tilde{\mathcal{F}}(f;N_p,\tau)) \Big)\bigg).
    \label{eq:CPMG_fit}
\end{align}
The empirical constant $C$ accounts for broadband noise reducing the overall contrast of the CPMG signal.

A CPMG measurement of line-synchronous noise was carried out with a string of eight ions.
A programmable RedPitaya~\cite{RedPitaya} board was used for generating a compensation current sent to the magnetic field coils of our experiment. The board was line-triggered for the creation of a line-synchronous feed-forward signal. To sense frequency components at 50\,Hz, 150\,Hz, and 250\,Hz, we chose a CPMG sequence with a duration of $\tau=20\,$ms with $N_p=2$, $N_p=6$, and $N_p=10$ $\pi$-pulses. Due to the fact that the contributions from the 150$\,$Hz and 250$\,$Hz components show up in the measurement for the 50$\,$Hz component, the compensation routine is started by compensating the higher frequency components first (before compensating the dominant 50\,Hz component). Towards this end, we fitted the measured signal by eq.~\eqref{eq:CPMG_fit} with $A$, $\phi_\mathrm{noise}$, and $C$ as fit parameters, and programmed the RedPitaya to apply an out-of-phase signal based on the calculated amplitude and phase to the magnetic field coils. The fine tuning is performed by trial-and-error and remeasuring the remaining signal.

Fig.~\ref{fig:CPMG_seq_mag_noise}(c) displays a CPMG measurement of the 50\,Hz component before compensation (red circles) and after (blue squares), together with fits shown as solid lines. The measured magnetic field strength of all three frequency components, before and after compensation, are listed in Table \ref{tab:mag_noise}. In general, a reduction of the modulation amplitude by at least a factor of ten was reached for all components. Fluctuations of the noise amplitudes and phases occurring on a time scale of minutes made a perfect compensation of line-synchronous noise impossible.\par

To test the effect of the feed-forward system on the qubit coherence, we carried out Ramsey contrast measurements as shown in Fig.~\ref{fig:CPMG_seq_mag_noise}(d) for a probe time of $\tau=4.5\,$ms. We compared the results of three cases; line trigger and feedforward compensation both ``on"  (red circles, $C=0.84(2)$), only the compensation ``on" (green triangles, $C=0.85(5)$), and both ``off"  (blue squares, $C=0.25(3)$). The strongly reduced contrast found in the latter case shows that in the absence of a feed-foward compensation, line-triggering the experiment is indispensable. Application of the feed-forward compensation restores the contrast and demonstrates that no additional noise is created by the feed-forward system. Using the feed-forward compensation, we do not have to rely on the line trigger anymore, which comes with the added benefit of a faster repetition rate of the experimental cycle.

\begin{table}[]
    \caption{Magnetic field strength before and after the compensation of the relevant noise components and the corresponding frequency shift $\Delta$ of the S$_{1/2} \leftrightarrow$ D$_{5/2}$ transition}
    \centering
    \begin{tabular}{|c|c|c|c|c|}
        \hline
        $f$ in Hz & $B$ in $\mu$G & $B$ after & $\Delta$ in Hz & $\Delta$ after\\ 
         & & compensation &  & compensation\\\hline
        50 & 37.2(5) & 1.3(6) & 104(2) & 3(2) \\
        150 & 9.3(8) & 0.9(6) & 26(2) & 2(2)\\
        250 & 23.3(6) & 0.7(5) & 65(2) & 2(1)\\\hline
    \end{tabular}
    \label{tab:mag_noise}
\end{table}

\subsection{\label{sec:Wavefront}  Wavefront tilt and curvature measurements}

Collective single-qubit operations as well as entangling operations are performed with a \SI{729}{nm} beam that is illuminating the entire ion string from a direction perpendicular to the ion string. Experimentally, we observe that a tilt in the wavefront of this transverse beam leads to infidelities of the collective single-qubit operations via a coupling to the comparatively hot axial modes of motion. Such a tilt can arise from an imperfectly aligned beam direction. Moreover, a spatially varying tilt can be caused by wavefront curvature, which can result, for example, if the ion string is situated outside the focus of a Gaussian beam. In this section, we show how such a wavefront tilt can be detected using the ions as a probe and how a proper beam shaping creates plane and parallel wavefronts which improves the fidelity of our collective qubit-operations.

In order to efficiently couple the available laser power to the linear ion string, we shape the transverse \SI{729}{nm} beam into an elliptical beam shape. Until recently, this was achieved with two crossed cylindrical lenses that created an astigmatic beam \cite{jurcevic2017}. The ion string was located at the vertical focus of beam radius of $24(6)\, \mu$m where the beam had a diameter of about $690(20)\,\mu$m in the horizontal direction. As a consequence of the ions being placed outside the horizontal focus, the ions were subjected to curved wavefronts. 

As a result, the quality of global qubit-operations was affected by spurious coupling to the string's axial modes of motion. This source of imperfections can be understood in terms of a simple semiclassical model: The axial COM-mode of the ion string is considerably populated even after polarization gradient cooling ($\bar{n}=$~170(20) at $\omega_z=2\pi \times$128~kHz). Therefore, each ion may be considered as a point-like particle oscillating back and forth with the axial trapping frequency. This ion oscillation leads to a phase modulation of the laser light seen by the ion if the wavefronts are not parallel to the ion string. The phase modulation gives rise to global unitaries with time-dependent rotation axes.

This effect can be observed in a Ramsey experiment. The ions are initially prepared in $\left| S \right>$ and a $\pi/2$-pulse is applied. After a wait time, a $\pi/2$-pulse with a phase shifted by $\pi$ ideally rotates the qubits back to the initial state. However, when scanning the wait time in such an experiment, we observe an excitation of up to 0.3\% that is modulated with the axial trapping period $T_\mathrm{ax}$. 

\begin{figure}
\centering
	\includegraphics[width=0.48\textwidth]{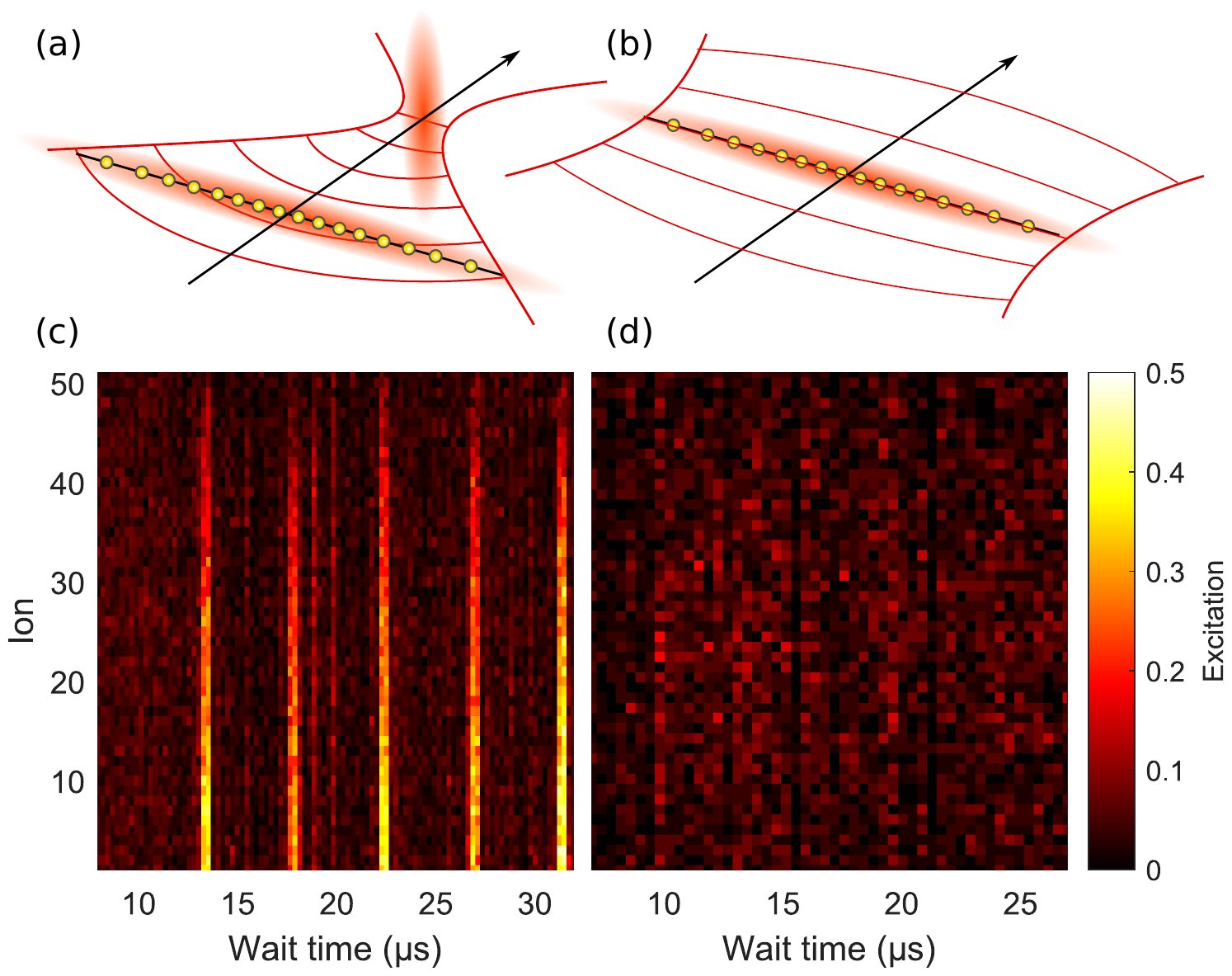}
	\caption{Wavefront detection using a CPMG sequence. The detected signal is produced by scanning the wait time between the pulses of the CPMG sequence. (a) Elliptical astigmatic beam (b) Elliptical non-astigmatic beam (c) The tilted and curved wavefronts of the astigmatic beam of the old beam shaping optics give rise to peaks under a CPMG sequence ($N_p$=20 pulses, axial trapping frequency $\omega_z=2\pi \times 112$~kHz, ion string temperature $T=4.6$~mK). (d) After the implementation of new beam shaping optics that deliver straight wavefronts, the peaks vanish.}
	\label{fig:CPMG_wavefronts}
\end{figure}

The coupling to the axial COM-mode can be amplified and detected more precisely by the use of a CPMG sequence, similar to the one described in sec. \ref{sec:LineCycle}. Again, the wait time between the $\pi$-pulses of the CPMG sequence is scanned. If the wait time equals $T_\mathrm{ax}\cdot(n+1/2)$, with $n \in \mathds{N}_0$, two consecutive $\pi$-pulses probe the phase modulation of the laser light at points of opposite sign which leads to an accumulation of the phase by the CPMG sequence. A sequence of 20 alternating $\pi$-pulses of length 4~$\mu$s gives rise to a series of distinct peaks separated by half a trapping period (Fig.~\ref{fig:CPMG_wavefronts}(a)). At one end of the ion string the peak excitation is $e_\mathrm{max} \approx 0.4$ and at the other end approximately zero. We attribute the variation of the peak height to a curvature of the transverse beam wavefronts hitting the ions in combination with a slight misalignment of the beam's k-vector. The excitation for a wait time of $T_\mathrm{ax}\cdot(n+1/2)$ can be estimated from the semiclassical model as (see appendix \ref{sec:APPENDIX_semiclassical})
\begin{equation}
    \label{eq:exc_thermal_peak_main}
    p_\mathrm{D} = e_\mathrm{max}
	= \frac{1}{2} \left(1 - \mathrm{e}^{-\frac{2k_\mathrm{B}T k_z^2 (N_p+1)^2}{m\omega_z^2} } \right),
\end{equation}
where $N_p$ is the number of $\pi$-pulses, $\omega_z$ the axial trapping frequency, $k_z$ the wave vector component along the axial trap axis, $T$ the ion temperature, $m$ the ion mass and $k_\mathrm{B}$ is Boltzmann's constant. For the experimental parameters used in the CPMG-experiment ($N_p=20$, $\omega_z=2\pi \times 112$~kHz, $T=4.6$~mK and $m=m_\mathrm{Ca40}$) a variation of the angles between the local wavefront and the axial direction from $\alpha_{1}\approx4.8$~mrad to $\alpha_{51}\approx1.4$~mrad over the ion string would give rise to the measured peak excitation. This variation of $\Delta \alpha = \SI{3.4}{mrad}$ over an ion string of length $\Delta z = 269~\mu$m corresponds to a wavefront curvature of radius $R=\Delta z / \Delta \alpha = \SI{79(2)}{mm}$. The ions were placed $\SI{36(2)}{mm}$ from the horizontal focus. Thus, the expected wavefront curvature was $\SI{36(2)}{mm}$, being significantly smaller than the value inferred from the wavefronts. A possible explanation for this discrepancy could be distortions in the wavefronts of the beam.

We improved the quantum operations induced by the global laser beam  by replacing the astigmatic beam with an elliptical, non-astigmatic beam. This beam is shaped by two crossed cylindrical-lens telescopes. The foci along the horizontal and the vertical axis coincide whereby plane wavefronts are created along both axes. At the ion string, the beam has a radius ($1/e^2$-intensity) of $235(10)$~$\mu$m and $23(4)$~$\mu$m, respectively. Probing the plane wavefronts with a CPMG sequence shows that the peaks vanish as soon as the wavefronts are made parallel with respect to the ion string (Fig.~\ref{fig:CPMG_wavefronts}(b)). While no peaks are visible in the CPMG signal, the noise floor gives an upper bound on the wavefront tilt of \SI{0.8}{mrad}.

Interestingly, the CPMG signal shows not only peaks that are separated by a full trapping period but also intermediate peaks that are separated by half a trapping period. While these peaks are not predicted by a semiclassical model assuming infinitely fast $\pi$-pulses, we find that a numerical solution of the Schrödinger equation describing the action of the CPMG sequence gives rise to intermediate peaks~(see App.~\ref{sec:APPENDIX_numerical}). The intermediate peaks get the strongest when the Rabi frequency of the light-matter interaction is of similar size as the trapping frequency. In the experimental data (Fig.~\ref{fig:CPMG_wavefronts}(c)) there is no intermediate peak visible at 9~$\mu$s for which we do not have an explanation so far.

\subsection{\label{sec:PathLengthFluctuations} Path-length fluctuations of optical beams}

Coherent laser-ion interactions are affected by phase noise that can occur from three main sources: phase noise from the laser itself, fluctuating electromagnetic fields shifting the atomic energy levels, or instabilities in the beam path from laser to ion. In the following, we will investigate the latter by designing a Ramsey-type experiment, with two different beam paths, that is sensitive only to relative path length fluctuations.

We consider a single ion addressed by light from two separate beam paths, termed 1 and 2. The light incident on the ions from each path has electric fields $\sim\text{e}^{i(\omega_{1}t_{1} +\phi_{1}(t))}$ and $\sim\text{e}^{i(\omega_{2}t_{2} +\phi_{2}(t))}$, respectively, where $\phi_i(t)$ accounts for laser phase noise and optical path length flucutuations between the laser and the ion. If it is assumed that any beam path fluctuations which may be present in the two paths are independent, the phases from the two separate beam paths will consequently be uncorrelated. Correlations in time between measurements of consecutive Ramsey experiments at times $i$ and $j$ then take the form \cite{Brydges:2021}:
\begin{equation}
\mathcal{C}_{i,j} = \langle\text{cos}(\Delta\phi_{i}-\Delta\phi_{j})\rangle,
\end{equation}
with $\Delta\phi_{i}=(\phi_{1}-\phi_{2})_{i}$. As such, the correlations between two such Ramsey experiments can be used to determine temporal changes of the phase difference between the first and second laser pulse. As this experiment is insensitive to laser phase noise for short Ramsey probe times, instabilities in the path lengths are the only contributions to the phase noise, and so the dynamics of these correlations will reveal information about fluctuations in the path lengths from one experiment to the next.

\begin{figure}
	\centering
	\includegraphics[scale=0.5]{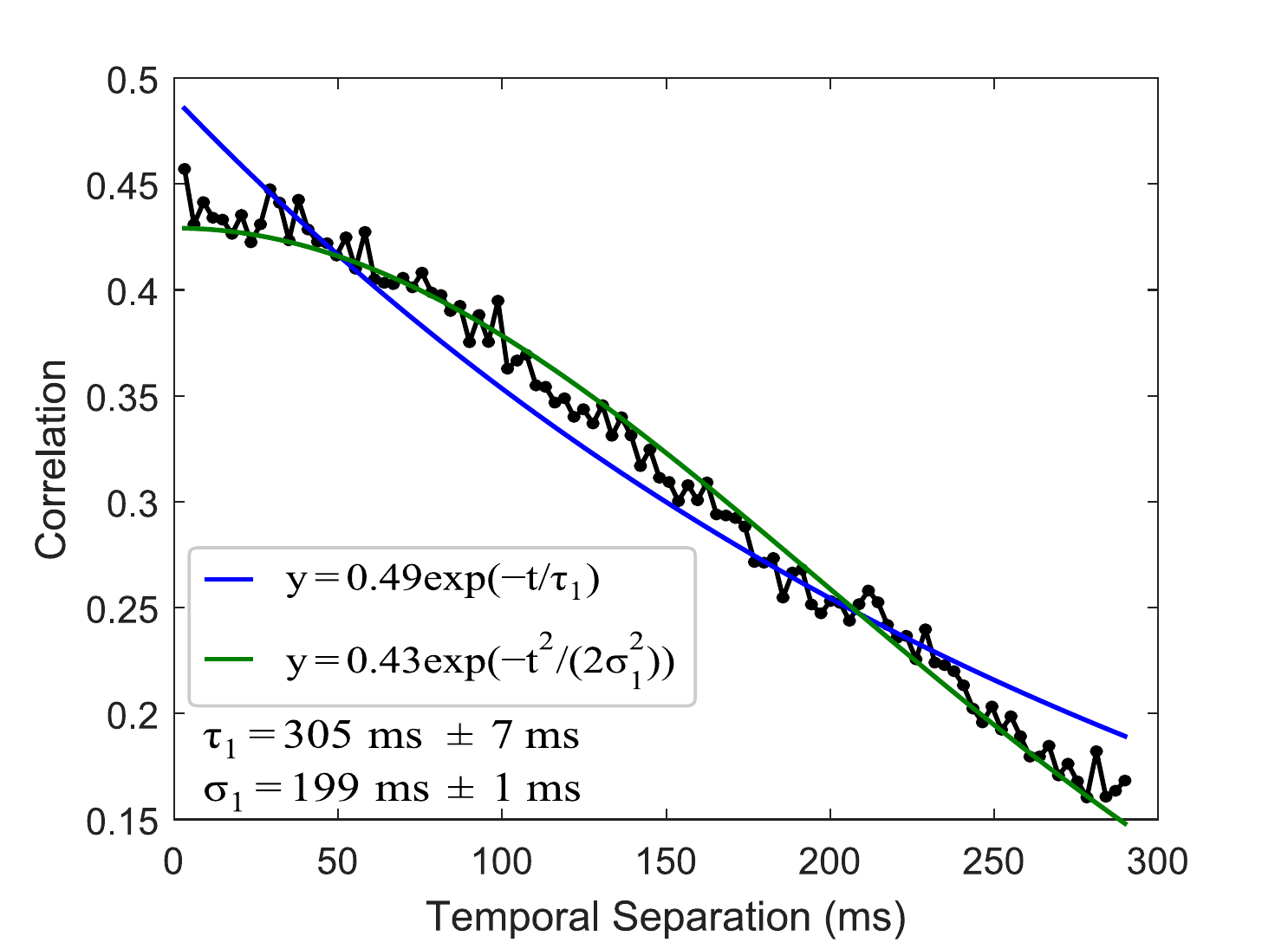}
	\caption{Auto-correlations from a single-ion experiment. Plotted are both exponential (blue line) and Gaussian (green line) fits to the data (black points). The Gaussian decay better fits the data, indicating that the decay in correlations is predominantly affected by slow phase noise.}
	\label{fig:singioncorrs}
\end{figure}

Practically, this experiment was implemented using the radial beam path and the addressing beam path (indicated in Fig.~\ref{fig:level_scheme_setup}(a)), and a short Ramsey probe time on the order of 5~$\mu$s, with the phase of the second pulse then scanned over. Fig.~\ref{fig:singioncorrs} shows the results from performing this experiment. The figure shows the correlations of the ion with itself as a function of the length of time separating the two measurements. Shown are two fits to the data: an exponential decay (blue line) and a Gaussian decay (green line). For fast frequency noise, an exponential decay of the correlations can be expected. For low-frequency noise, the decay-curve is likely to be non-exponential, with a Gaussian shape expected \cite{home:2006}. It can be seen that a Gaussian decay better fits the decay of these correlations over the short time-scales probed in the experiment, indicating that the correlations here are predominantly affected by slow phase noise. This is consistent with phase noise produced by fluctuations in the beam path, which would be expected to be slow drifts.

The correlations decay over a time-scale of approximately \SI{300}{ms}, which is much larger than the current coherence time of the system (on the order of \SI{64}{ms} for the $|\text{S}\rangle$ to $|\text{D}\rangle$ transition \cite{Brydges:2021}), and so it can be concluded that phase instabilities introduced by path length fluctuations are not, currently, a dominant source of noise in the system.

\section{\label{sec:Addressing} Individual quantum state control of long ion strings}
\begin{figure}
\centering
\includegraphics[width=0.48\textwidth]{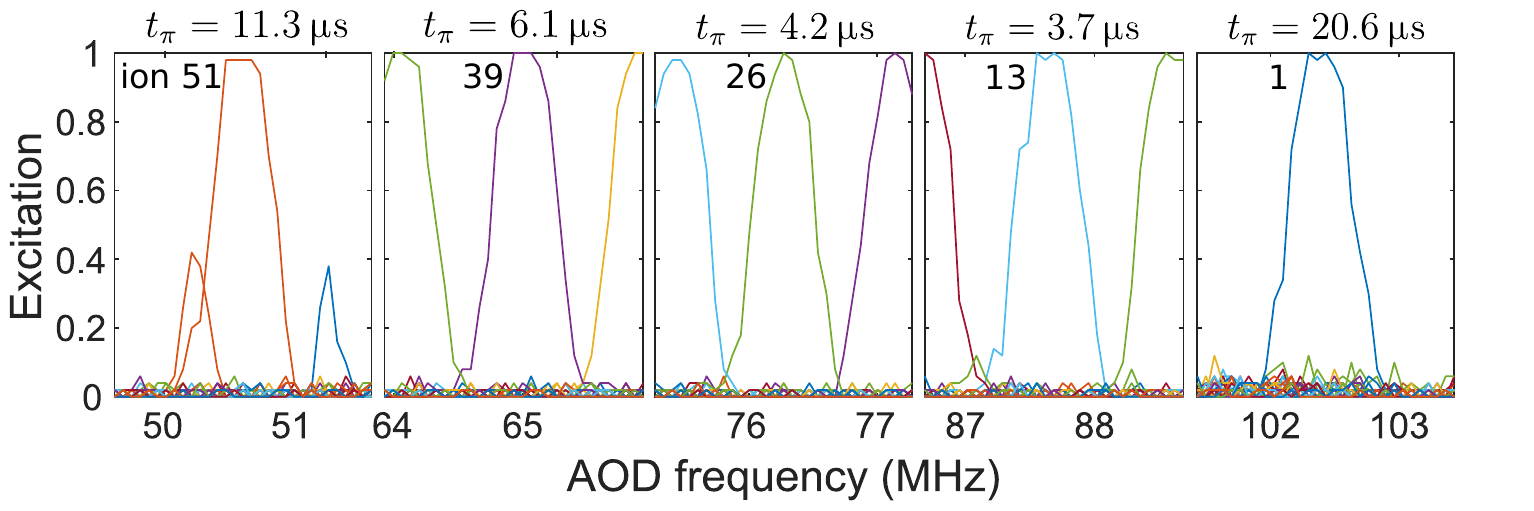}
\caption{Single-ion addressing of a 51-ion string illustrated for a few selected ions. The ions were initialized in the $X$-basis, then addressed by a laser pulse inducing an AC-Stark shift with the pulse length set to realize a $\pi$-rotation, and measured in the $X$-basis. For ion 51 the double-frequency component of the AOD leads to an excitation of ion 1 and 2.}
\label{fig:51_addressing}
\end{figure}

We perform single-qubit gate operations with a tightly focused laser beam at \SI{729}{nm} whose focal spot is steered over the ion string using an acousto-optic deflector (AOD)~\cite{AOD}. A system of five lenses maps the output of the AOD onto an objective, such that each beam defracted by the AOD is projected to the same area on the objective. The objective focuses the beam to about 2~$\mu$m radius.
With this setup, we achieve individual ion addressing in a 51-ion string, which, at an axial confinement frequency $\omega_z =2\pi \times 127$~kHz, extends over a range of 246 $\mu$m. Addressing of the ions 1, 13, 26, 39 and 51 is demonstrated in Fig.~\ref{fig:51_addressing}, which shows the excitation probability of the individual ions as a function of the AOD frequency. 

\begin{figure}
\centering
\includegraphics[width=0.45\textwidth]{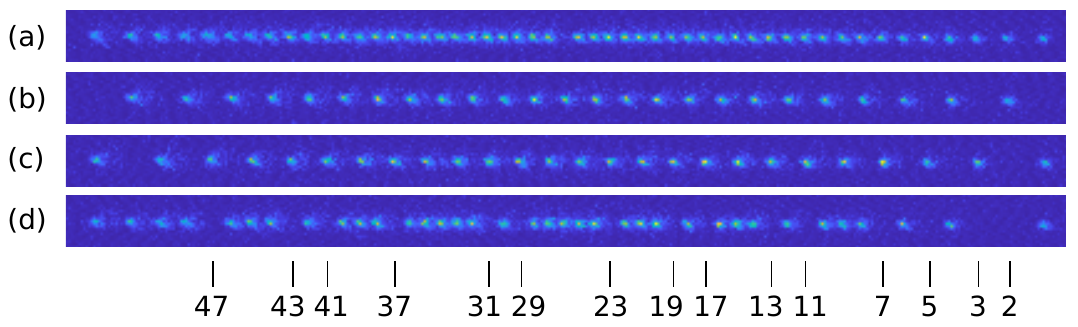}
\caption{Ions prepared in different initial states: (a) Ion 26 addressed, (b) Néel-state with odd ions addressed, (c) Néel-state with even ions addressed, and (d) ions that have a prime number as index addressed (ions numbered from right to left).}
\label{fig:variousInitStates}
\end{figure}

Single-qubit gates are realized by using the strongly focused beam for inducing AC-Stark shifts on individual qubits in combination with laser pulses from the global beam that induce resonant $\pi/2$ pulses on all ions simultaneously. By sandwiching AC-Stark-shift induced $\pi$-pulses on a sub-ensemble of ions between a pair of resonant $\pi/2$-pulses of opposite phase, product states can be produced, in which the qubits are initialized in the up/down state, as shown for various examples in  Fig.~\ref{fig:variousInitStates}. More generally, it is possible to induce arbitrary (ion-specific) single-qubit operations on all ions simultaneously \cite{Brydges:2019}. Among others, this capability enables measurements of arbitrary correlation functions between the qubits as required for quantum state tomography.

\begin{figure}
\centering
\includegraphics[width=0.45\textwidth]{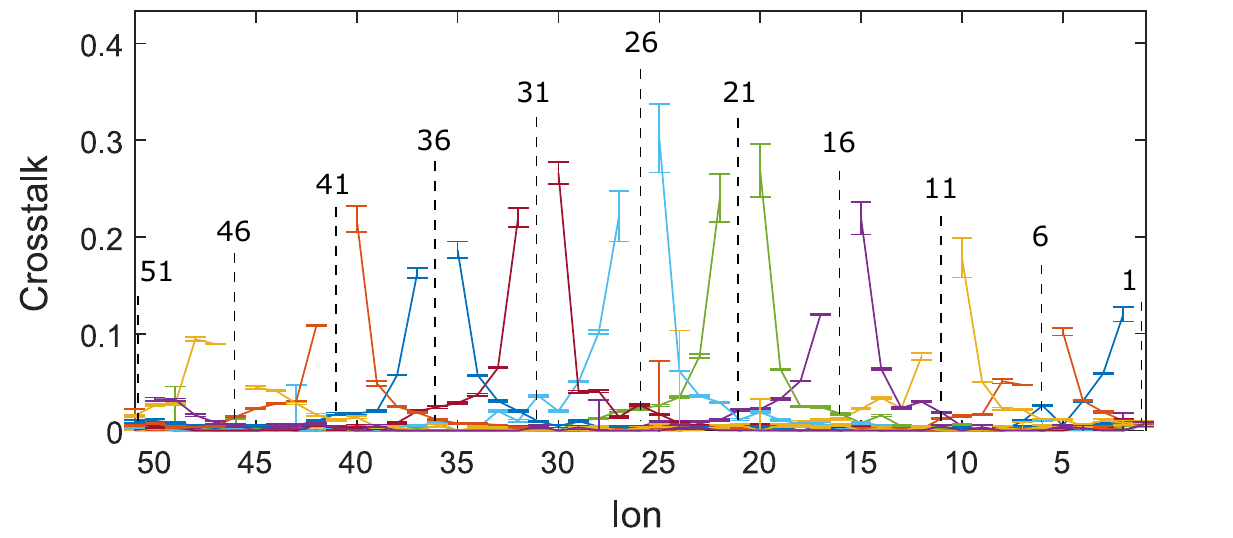}
\caption{Cross-talk in a 51-ion string, measured as the ratio of the Rabi-frequency of a non-addressed and of the addressed ion. This measurement was repeated for every fifth ion in the ion string.}
\label{fig:51_addressingCrosstalk}
\end{figure}

We characterize the addressing performance of our system by a cross-talk analysis. Due to the finite beam size, the neighboring ions of the ion of interest are also illuminated by some laser light, which coherently rotates the qubit state, thus causing an unwanted operation. In the case of a resonant addressing beam, the neighboring ions experience a cross-talk of 0.03-0.3 (ratio of Rabi frequencies), depending on the position of the addressed ion in the string (see Fig.~\ref{fig:51_addressingCrosstalk}). The average next-neighbor cross-talk per qubit is reduced to $3 \times 10^{-2}$ (ratio of intensities) by performing AC-Stark gates instead of resonant operations for the single-qubit gates.

\begin{figure}
\centering
\includegraphics[width=0.45\textwidth]{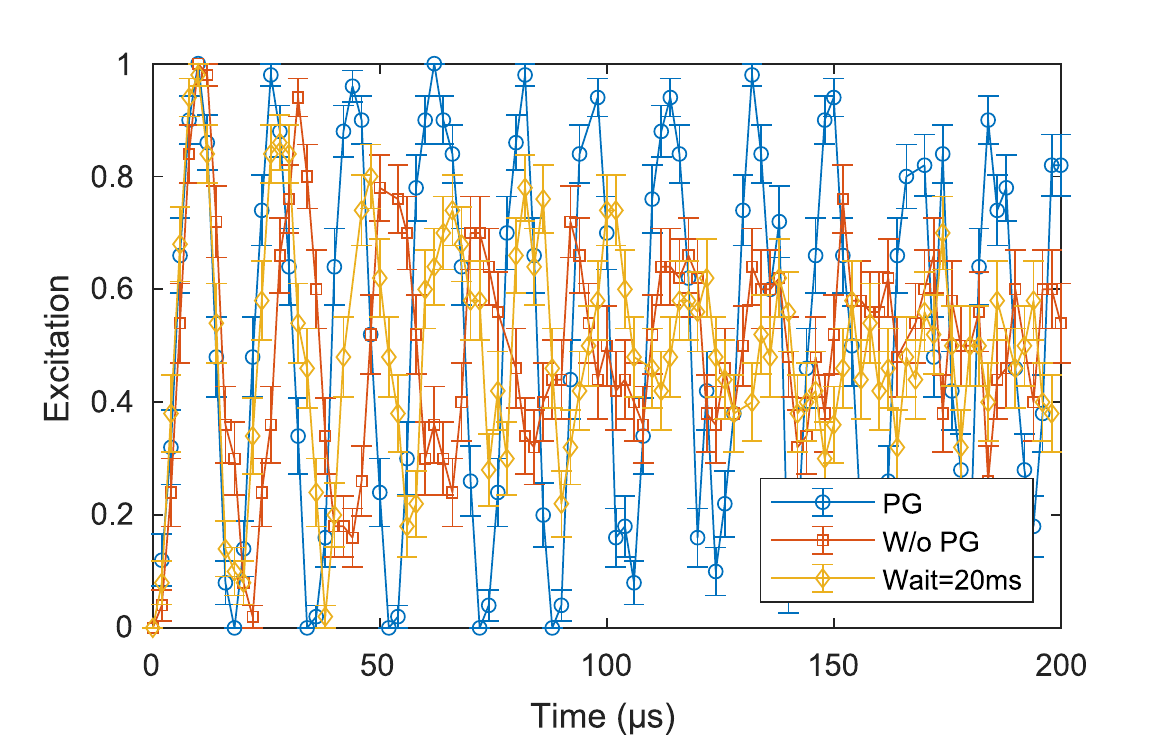}
\caption{Addressing flop quality after PG cooling for 1 ms (blue) without PG cooling (red) and with an extra 20 ms waiting time after PG cooling (orange) of a 51 ion string. Here we show only addressing flops for the 26$^{th}$ ion. Solid lines are a guide to the eye. }
\label{fig:Addr_vs_time}
\end{figure}

Motional heating of the axial COM mode compromises the quality of our single-qubit gate operations, as illustrated in  Fig.~\ref{fig:Addr_vs_time}. To reduce this effect, we cool the axial motional modes before and after sideband cooling of the radial modes, via the polarization-gradient cooling technique; still, motional heating remains a source of imperfections for long experimental sequences.

Another effect that is relevant for single-ion addressing is drift of the ion positions in the weakly confining  direction. We observe drifts of the positions by about \SI{40}{nm} along the direction of the string within a few minutes. These drifts may be caused by surface charging effects triggered
by the Doppler cooling beam (\SI{397}{nm}). To counter this drift, we automatically detect the ion positions every 5-10 minutes using the EMCCD camera and cancel the shift by feeding back onto the voltage of the endcap electrodes. Subsequently, we automatically recalibrate the addressing Rabi frequency of each ion. 

\section{Entanglement generation } \label{sec:entanglement}
As discussed in Sec. \ref{sec:setup}, we realize a M{\o}lmer-S{\o}rensen type interaction whose effective Hamiltonian is the one of a long-range XY model. For this, a large centerline detuning has to be chosen ($\delta=2\pi \times 3000$~Hz) as compared to the elements of the $J_{ij}$ matrix (i.e. $ \le 240$ rad/s). To observe the spin dynamics induced by such an interaction, we prepare a  N\'eel state, i.e. a state with alternating spin orientations, in the $z$-basis, and time-evolve the initial state under the entangling interaction. Measurements of the time-evolved state in the $z$ basis are shown in Fig.~\ref{fig:EntaglementGen}(a). 

\begin{figure}
    \centering
   \includegraphics[width=0.45\textwidth]{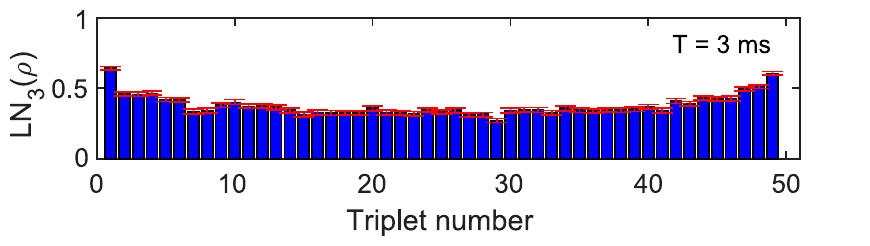}
    \caption{Log-negativities measured for adjacent ion triplets after $T=\SI{3}{ms}$ of entangling operation in a 51-ion chain (for the corresponding plot showing log-negativities of pairs, please see Fig.~\ref{fig:EntaglementGen}(b)).}
    \label{fig:LogN51}
\end{figure}

The entanglement generated by the dynamics is assessed by tomographic measurements of reduced 3-qubit density matrices, enabling the calculation of the logarithmic negativity for adjacent ion pairs and triplets \cite{lanyon2017}. For pairs and triplets, the logarithmic negativity is expressed \cite{lanyon2017} as 
\begin{eqnarray}
    \text{LN}_2(\rho) &=& \log_2 ||\rho^{T_A}||_i,\\
    \text{LN}_3(\rho) &=& \sqrt[3]{\text{LN}_2(\rho_{i,jk})\text{LN}_2(\rho_{j,ik})\text{LN}_2(\rho_{k,ij})},
\end{eqnarray}
where LN$_2(\rho)$ is calculated by taking the logarithm of the partial transpose of a two-qubit density matrix over the $i^\mathrm{th}$ qubit for a pair formed by $i^\mathrm{th}$ and $(i+1)^\mathrm{th}$ ions. On the other hand, LN$_3(\rho)$ is a geometrical mean of three connected, two-qubit logarithmic negativities. The results are presented in Fig.~\ref{fig:EntaglementGen}(b) for adjacent ion pairs and in 
Fig.~\ref{fig:LogN51}  for triplets, for the case of a N\'eel state time-evolved for $T=3$ ms. At this time, all pairs and triplets of adjacent ions turn out to be entangled, whereas at longer times larger subsystems would have to be analyzed in order to detect entanglement. 

\section{\label{sec:Conclusions} Conclusions} 

We have presented methods for controlling long ion strings for quantum simulation and precision experiments. We presented heating rate measurements for ion strings of up to 51 ions. The time span that is available for experiments with the ion string is limited by collisions with the background gas and chemical reactions. For a 51-ion string we found a life time of 27.0(2) s until the ion string melts after a collision. For cooling the motional modes of long ion strings that are trapped in a highly anisotropic trapping potential we use Doppler cooling, sideband cooling and polarization-gradient cooling methods.

For enabling coherent manipulation of long qubit registers we have demonstrated several techniques: A feed-forward scheme was applied to compensate line-synchronous magnetic field components. The magnetic field variations for these frequency components have been reduced by more than an order of magnitude in our apparatus. Auto-correlation measurements performed with a single ion were used to measure path-length fluctuations, which were shown to not be a limiting factor for the coherence time of our system. We have demonstrated how the wavefront curvature of the global beam can be measured using the ion string as a probe and how providing plane wavefronts improves the fidelity of global qubit-operations. We have demonstrated the individual-ion addressing capability of our experimental platform. The full ion string of 51 ions can be prepared in arbitrary product states. We showed how the addressing fidelity benefits from polarization-gradient cooling. Finally, by utilizing individual qubit control and entangling operations, the entanglement generation is verified for 51-ion strings by detecting entanglement for all connected pair and triplets.

With the work presented in this paper we gain the capability to perform quantum simulations with large, fully individual-ion addressable systems of 51 ions~\cite{joshi2021hydro}. Such large systems~\cite{zhang2017observation, bernien2017probing, browaeys2020many, arute2019quantum} hold the promise to study physics that is inaccessible by numerical computation. 

\begin{acknowledgements}
The project leading to this application has received funding from the European Union’s Horizon 2020 research and innovation programme under grant agreement No 817482. Furthermore, we acknowledge support by the Austrian Science Fund through the SFB BeyondC (F7110) and funding by the Institut f\"ur Quanteninformation GmbH.
\end{acknowledgements}

\appendix
\section{\label{sec:APPENDIX} Modelling wavefront curvature measurements by CPMG sequences}

\subsection{\label{sec:APPENDIX_semiclassical}Derivation of the semiclassical model}

The excitation of a hot ion after after application of a CPMG pulse sequence can be approximately calculated by modelling the ion as a point-like particle undergoing classical harmonic motion and assuming $\pi$-pulses of a duration much shorter than the trapping period. The derivation is done in two steps: First, the excitation after a $\pi$-pulse sequence is calculated for a given trajectory. Second, the mean excitation is averaged over a thermal distribution of the axial COM-mode.

Initially, the ion is prepared in $\left| \downarrow \right> = \left| \mathrm{S} \right>$. Between an initial and a final $\pi/2$-pulse a series of $N_p$ alternating $\pi$-pulses is applied. The phase of each pulse is determined by the position of the ion in the light field driving the pulses. The light-matter interaction is described in the basis $\begin{pmatrix} 1 \\ 0 \end{pmatrix} = \left| \uparrow \right> = \left| \mathrm{D} \right>$ and $\begin{pmatrix} 0 \\ 1 \end{pmatrix} = \left| \downarrow \right> = \left| \mathrm{S} \right>$ by the unitary operation
\begin{equation}
	\label{eq:rotation_general}
	U(\theta, \phi) = \begin{pmatrix}
		\cos \frac{\theta}{2} & -\mathrm{i} \mathrm{e}^{-\mathrm{i}\phi}\sin{\frac{\theta}{2}} \\
		-\mathrm{i} \mathrm{e}^{\mathrm{i}\phi}\sin{\frac{\theta}{2}} & \cos \frac{\theta}{2} \\
	\end{pmatrix},
\end{equation}
where $\theta$ denotes the rotation angle and $\phi$ the angle of the rotation axis within the Bloch sphere's equatorial plane. The total pulse sequence is described by
\begin{align}
	U = U(\pi/2, \pi/2 + \phi_\mathrm{f}) 
		U((-1)^{N_p+1} \pi, \phi_{N_p}) \cdots \\ \nonumber
		U(-\pi, \phi_2) U(\pi, \phi_1)
		U(\pi/2, 3\pi/2 + \phi_\mathrm{i}), 
\end{align}
where $\phi_\mathrm{i}$, $\phi_\mathrm{f}$ and $\phi_1, \dots, \phi_{N_p}$ denote the change of the angle of the rotation axis at the time when the respective pulse is applied. In the limit of vanishing motional excitation, we assume the $\pi/2$-pulses to be described by rotations around the $\pm y$-axis and the $\pi$-pulses by rotations around the $\pm x$-axis. The excitation after the pulse sequence is
\begin{equation}
	\label{eq:excitation_general}
	e = p_{\uparrow} = |\left< \uparrow | U | \downarrow \right>|^2 = \frac{1}{2} \left(
			1 - \cos \Phi
		\right),
\end{equation}
where the accumulated phase
\begin{equation}
	\Phi = \phi_\mathrm{i} + 2\sum_{n=1}^{N_p} (-1)^{n} \phi_n + (-1)^{N_p+1}\phi_\mathrm{f}
\end{equation}
has been introduced.

For the case of temporally equidistant $\pi$-pulses separated by a wait time $T_\mathrm{wait}$ acting on a harmonically trapped ion with trapping frequency $\omega$ the accumulated phase takes a simple form. The times at which the pulses occur are $t_\mathrm{i} \in \mathds{R}$, $t_n = t_\mathrm{i}+nT_\mathrm{wait}$ for $n=1,\dots,N_p$ and $t_\mathrm{f} = t_\mathrm{i} + (N_p+1)T_\mathrm{wait}$. The classical ion trajectory of amplitude $a$,
\begin{equation}
	z(t) = a \sin(\omega t)
\end{equation}
then gives the phases for the pulses
\begin{align}
	\label{eq:initial_phase}
	&\phi_\mathrm{i}(t_\mathrm{i})=k_z z(t_\mathrm{i})=
		k_z a \sin(\omega t_\mathrm{i}),\\
	&\phi_\mathrm{n}(t_\mathrm{i})=
		k_z a \sin(\omega (t_\mathrm{i}+nT_\mathrm{wait})),\quad \mathrm{for~} n=1,\dots,N_p,\\
	\label{eq:final_phase}
	&\phi_\mathrm{f}(t_\mathrm{i})=
		k_z a \sin(\omega (t_\mathrm{i}+(N_p+1)T_\mathrm{wait})),
\end{align}
where $k_z$ is the component of the wave vector along the ion trajectory.
For this case the accumulated phase $\Phi$ can be written as
\begin{equation}
	\label{eq:accumulated_phase_harm}
	\Phi = k_z a \left[ 
		\sin(\omega t_\mathrm{i}) A_{N_p}(\omega T_\mathrm{wait}) + 
		\cos(\omega t_\mathrm{i}) B_{N_p}(\omega T_\mathrm{wait})
	\right],
\end{equation}
where the abbreviations
\begin{align}
    A_{N_p} (\omega T_\mathrm{wait}) & =  1 + 2 \sum_{n=1}^{N_p} (-1)^{n} \cos(n\omega T_\mathrm{wait}) \nonumber \\
	& + (-1)^{N_p+1} \cos((N_p+1)\omega T_\mathrm{wait}),\\
	B_{N_p} (\omega T_\mathrm{wait}) & =  2 \sum_{n=1}^{N_p} (-1)^{n} \sin(n\omega T_\mathrm{wait}) \nonumber \\
	& + (-1)^{N_p+1} \sin((N_p+1)\omega T_\mathrm{wait})
\end{align}
are introduced. Using the accumulated phase $\Phi$ (eq.~\eqref{eq:accumulated_phase_harm}) the excitation can be directly evaluated with equation~\eqref{eq:excitation_general}. Here, the phase of the initial pulse with respect to the ion oscillation has to be known and the ion has to have a fixed energy.

In a realistic scenario, the excitation is the result of averaging over multiple repetitions of the experiment. Since the ion is in a thermal state neither the phase of the initial pulse nor the energy is the same between repetitions. Therefore, the expectation value over the thermal distribution has to be calculated. 

The thermal distribution is modeled by a Boltzmann distribution which gives the excitation
\begin{equation}
	e = \left< p_\uparrow(q,p) \right>
	= \int \mathrm{d}\Gamma Z^{-1} \mathrm{e}^{-\beta H(q,p)} p_\uparrow(q,p),
\end{equation}
where $\mathrm{d}\Gamma = \mathrm{d}q \mathrm{d}p$, $\beta=1/k_\mathrm{B}T$, $Z = \int \mathrm{d}\Gamma \mathrm{e}^{-\beta H(q,p)}$ denotes the partition sum and
\begin{equation}
	H(q,p) = \frac{1}{2} m\omega^2 q^2 + \frac{p^2}{2m}.
\end{equation} 
In order to average over the thermal distribution, the outcome of the experiment has to be expressed in terms of the initial point $(q,p)$ that is picked from phase space. Since the time-evolution for this initial point is simply the oscillation of a harmonic oscillator, its trajectory is described by $q(t) = a \sin(\omega t + \delta)$ with some phase $\delta$. The phases at which the $\pi$-pulses occur are described by equations \eqref{eq:initial_phase}-\eqref{eq:final_phase}. The accumulated phase after the $\pi$-pulse sequence can be expressed in terms of the initial coordinates in phase space $(q,p)$ by using
\begin{align}
	&q = a \sin(\omega t + \delta),\\
	&p = m \omega a \cos(\omega t + \delta)
\end{align}
which yields
\begin{equation}
	\Phi(q,p) = k_z A_{N_p} q + k_z \frac{B_{N_p}}{m \omega} p.
\end{equation}
The excitation is then
\begin{align}
	e = \int \mathrm{d}\Gamma Z^{-1} \mathrm{e}^{-\beta H(q,p)} \frac{1}{2} \left[ 1 - \cos(\Phi(q,p)) \right].
\end{align}
The partition sum is
\begin{equation}
	\label{eq:partition_sum}
	Z = \int \mathrm{d}\Gamma \mathrm{e}^{-\beta H(q,p)} 
	= \frac{2\pi}{\beta} \frac{1}{\omega}	.
\end{equation}
The second part of the integral consists of evaluating the expectation value of the cosine,
\begin{eqnarray}
	\left< \cos(\Phi(q,p)) \right>
	&= Z^{-1} \int \mathrm{d}\Gamma \mathrm{e}^{-\beta \left(\frac{1}{2} m \omega^2 q^2 + \frac{p^2}{2m} \right)} \nonumber \\ 
	&\times \cos\left(k_z A_{N_p} q + k_z \frac{B_{N_p}}{m \omega} p
	\right) \nonumber \\ 
	&= \mathrm{e}^{-\frac{k_\mathrm{B}T}{2m\omega^2} k_z^2 (A_{N_p}^2+B_{N_p}^2)}.
 \end{eqnarray}
Finally, we can write the excitation of an ion after a $\pi$-pulse sequence as
\begin{eqnarray}
	\label{eq:exc_thermal}
	e & = &  \frac{1}{2} - \frac{1}{2}\left< \cos(\Phi(q,p)) \right> \nonumber \\ 
	& = & \frac{1}{2} \left( 1 - \mathrm{e}^{-\frac{k_\mathrm{B}T}{2m\omega^2} k_z^2 C_{N_p}^2} \right),
\end{eqnarray}
where $C_{N_p}^2:=A_{N_p}^2+B_{N_p}^2$ has been introduced. A simple but lengthy calculation shows that
\begin{equation}
    C_{N_p}^2 = 4 \left( \frac{\sin \big((N_p+1)\frac{1}{2}(\omega T_\mathrm{wait} + \pi) \big)}{\tan \big(\frac{1}{2}(\omega T_\mathrm{wait} + \pi) \big)} \right)^2.
\end{equation}
The excitation reaches its maximum when $\omega T_\mathrm{wait}$ is an odd integer multiple of $\pi$, $\omega T_\mathrm{wait} = 2(n+1)\pi$, with $n \in \mathds{Z}$. At these points, $C_{N_p}^2$ takes the maximum value $C_{N_p,\mathrm{max}}^2 = 4(N_p+1)^2$.  This gives the peak excitation
\begin{equation}
\label{eq:exc_thermal_peak}
	e_\mathrm{max}
	= \frac{1}{2} \left( 1 - \mathrm{e}^{-\frac{2k_\mathrm{B}T k_z^2 (N_p+1)^2}{m\omega^2} } \right).
\end{equation}

\subsection{\label{sec:APPENDIX_numerical}Numerical solution of the quantum case}
The simple semiclassical model can explain the main peaks of the observed CPMG signal; however, it fails to explain intermediate peaks. To explain these intermediate peaks, too, the Schrödinger equation is solved numerically. The simulation is carried out using QuTiP~\cite{johansson2013qutip} based on the following steps: (1) The initial state is assigned to the qubit. (2) The Schrödinger equation for the Hamiltonian~\eqref{eq:lm-interaction} is solved for each part of the temporal evolution of the CPMG sequence consecutively. (3) The excitation probability $p_\uparrow$ is extracted from the final state.

In order to keep the problem in a numerically accessible regime, the light-matter interaction is modeled as a qubit moving in a one-dimensional harmonic potential and interacting with a travelling-wave laser beam. The Hamiltonian in the interaction picture is
\begin{equation}
    \label{eq:lm-interaction}
    H_\mathrm{I} = \hbar \omega \left( a^\dag a + \frac{1}{2} \right) + \frac{1}{2} \hbar \Omega \left(
    \mathrm{e}^{\mathrm{i} \eta (a+a^\dag)} \sigma^+ \mathrm{e}^{-\mathrm{i} \Delta t} + h.c.
    \right),
\end{equation}
where $a$ is the annihilation operator, $\sigma^+$ the raising operator, $\Omega$ the Rabi frequency, $\Delta$ the laser detuning from the atom transition frequency, $\eta$ the Lamb-Dicke parameter and $t$ the time. The wavefront tilt angle $\alpha$ (measured against the perpendicular beam incidence) is taken into account by the Lamb-Dicke parameter, $\eta = \cos(\pi/2-\alpha) \sqrt{\hbar/(2 m\omega)}$.

To calculate the CPMG signal of a thermal state, the problem is split up into two parts: First, the excitation $e_n$ is calculated for a set of Fock states $\left| \downarrow, n \right>$ and then averaged over a thermal distribution with mean phonon number $\bar n$.

\begin{figure}
   \centering
   \includegraphics[width=0.35\textwidth]{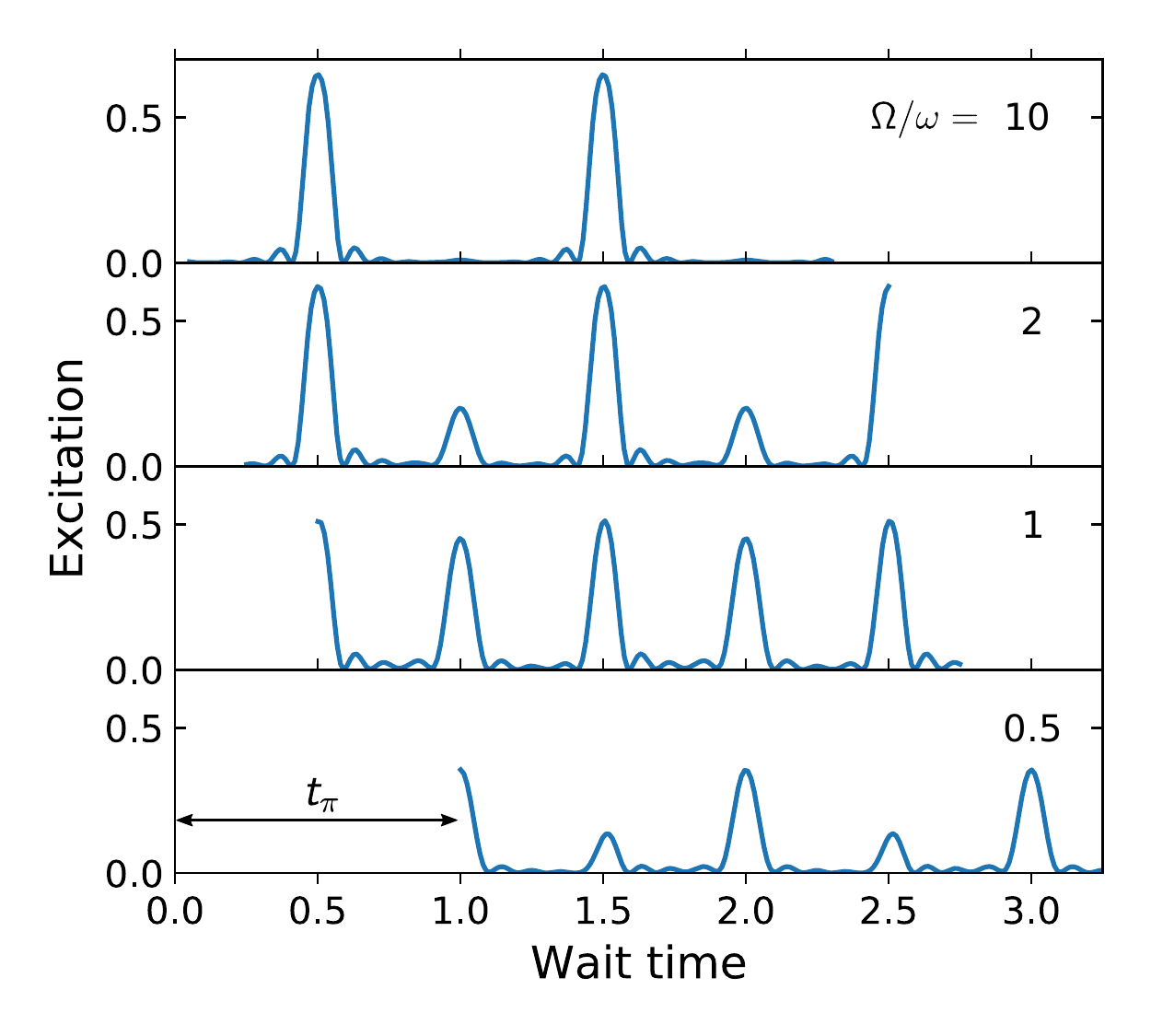}
    \caption{Excitation after a CPMG-10 sequence. The example of a numerical simulation for a Fock state $\left| \downarrow, n=50 \right>$ was calculated for the trapping frequency $\omega=2\pi$, Rabi frequency $\Omega$, $\Delta=0$ and $\eta=0.01$. The variation of the ratio $\Omega/\omega$ shows the emergence of the intermediate peaks for small Rabi frequencies, respectively large $\pi$-times $t_\pi=\pi/\Omega$. The wait time denotes the time difference between the start of two consecutive $\pi$-pulses.}
    \label{fig:CPMG_wavefronts_Rabi}
\end{figure}

\begin{figure}
   \centering
   \includegraphics[width=0.48\textwidth]{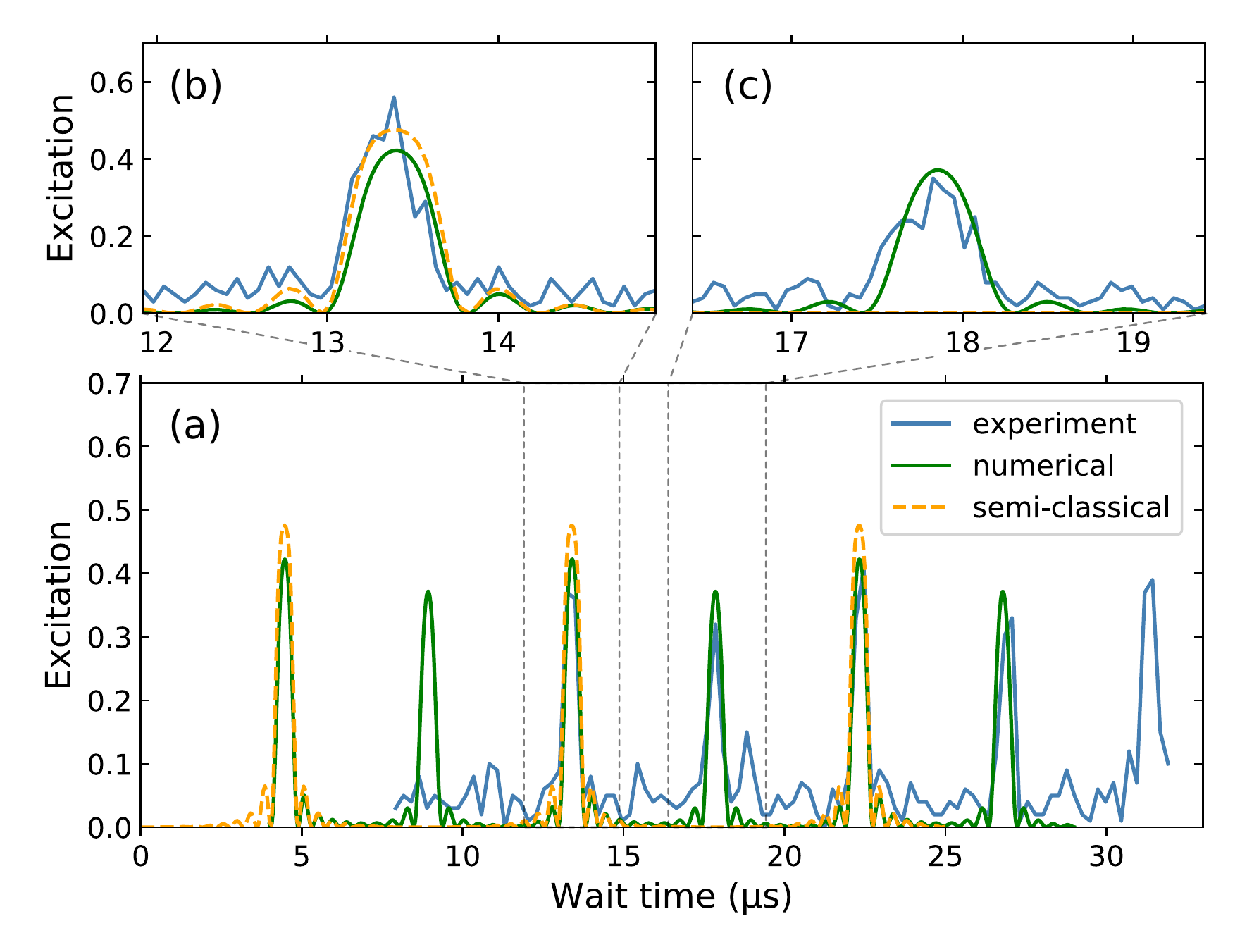}
    \caption{Comparison of numerical simulation and experiment. (a) The numerical simulation describes the full measured CPMG signal (ion 1), while the semiclassical model captures only the main peaks. (b,c) Detailed scans of the first and second peak show the different peak heights of the main and the indermediate peaks. The parameters were the same as for Fig.~\ref{fig:CPMG_wavefronts} ($N_p=20$ pulses, axial trapping frequency $\omega_z=2\pi \times 112$~kHz, ion string temperature $T=4.6$~mK).}
    \label{fig:CPMG_wavefronts_exp_num}
\end{figure}

The emergence of intermediate peaks, which are not predicted by the semiclassical model, are shown in Fig.~\ref{fig:CPMG_wavefronts_Rabi}. For a Rabi frequency that is much higher than the trapping frequency, the excitation shows peaks at $T_\mathrm{ax} (n+\frac{1}{2})$. As soon as the Rabi frequency becomes of the same order as the trapping frequency, intermediate peaks start to appear. A comparison between the experimental data and the numerical simulation shows a good agreement (Fig.~\ref{fig:CPMG_wavefronts_exp_num}).


%

\end{document}